\documentclass[sigconf]{acmart}

\usepackage{xcolor}
\usepackage{xspace}
\usepackage{enumitem}
\usepackage{subcaption}
\usepackage{tabularray}
\usepackage{listings}
\setlist[itemize]{left=0pt, labelindent=0pt}
\UseTblrLibrary{booktabs}
\AtBeginDocument{%
  }

\DeclareRobustCommand{\systemname}{\textbf{\textsc{Agent A/B}}\xspace}
\DeclareRobustCommand{\revise}[1]{#1}

\copyrightyear{2026}
\acmYear{2026}
\setcopyright{cc}
\setcctype{by}
\acmConference[CHI EA '26]{Extended Abstracts of the 2026 CHI Conference on Human Factors in Computing Systems}{April 13--17, 2026}{Barcelona, Spain}
\acmBooktitle{Extended Abstracts of the 2026 CHI Conference on Human Factors in Computing Systems (CHI EA '26), April 13--17, 2026, Barcelona, Spain}
\acmDOI{10.1145/3772363.3799039}
\acmISBN{979-8-4007-2281-3/2026/04}

\begin{document}

\title{\systemname: Automated and Scalable A/B Testing on Live Websites with Interactive LLM Agents}

\author{Yuxuan Lu}
\authornote{Work was done while interning at Amazon}
\affiliation{
    \institution{Northeastern University}
    \city{Boston}
    \state{Massachusetts}
    \country{USA}
}

\author{Ting-Yao Hsu}
\authornotemark[1]
\affiliation{
    \institution{Pennsylvania State University}
    \city{State College}
    \state{Pennsylvania}
    \country{USA}
}

\author{Hansu Gu}
\author{Limeng Cui}
\author{Yaochen Xie}
\author{William Headden}
\affiliation{
    \institution{Amazon}
    \country{USA}
}

\author{Bingsheng Yao}
\affiliation{
    \institution{Northeastern University}
    \city{Boston}
    \state{Massachusetts}
    \country{USA}
}

\author{Akash Veeragouni}
\author{Jiapeng Liu}
\author{Sreyashi Nag}
\author{Jessie Wang}
\affiliation{
    \institution{Amazon}
    \country{USA}
}

\author{Dakuo Wang}
\authornote{Contact: d.wang@northeastern.edu}
\affiliation{
    \institution{Northeastern University}
    \city{Boston}
    \state{Massachusetts}
    \country{USA}
}

\renewcommand{\shortauthors}{Lu et al.}

\newcommand{\kenneth}[1]{}
\newcommand{\edward}[1]{}
\newcommand{\cy}[1]{}
\newcommand{\limeng}[1]{}

\begin{abstract}

A/B testing is central to UI/UX design, yet our formative study with six industry practitioners revealed that it is slowed by scarce user traffic, long runtimes, and high operational costs.
To address these challenges, we introduce \textbf{\systemname}, an end-to-end system that deploys large language model (LLM) agents with structured personas to interact with live webpages and generate scalable behavioral evidence before launch. 
In a case study on Amazon.com, \systemname simulated a between-subjects A/B test of filter panel designs with 1,000 agents and found that the reduced filter list produced more purchases, reproducing directional outcomes also observed in a parallel large-scale human experiment. Results further suggest that agent-based simulations can detect interface-sensitive behavioral differences and surface subgroup patterns while offering faster, lower-risk insights. We position \systemname as a complement to human testing, enabling earlier prototyping, pre-deployment validation, and hypothesis-driven UX evaluation.

\end{abstract}


\begin{teaserfigure}
  \includegraphics[width=\textwidth]{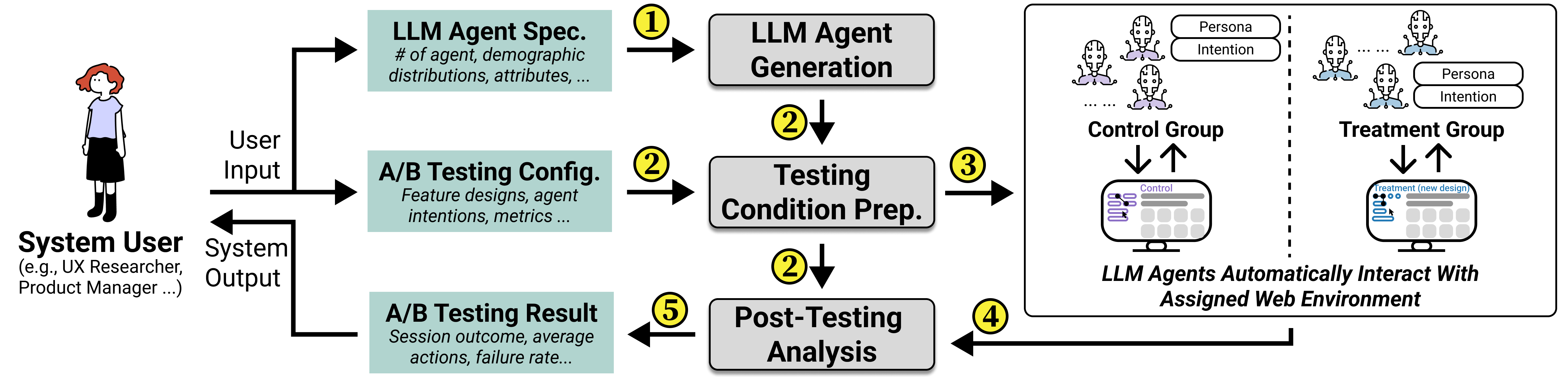}
  \caption{%
  Architecture of \systemname. From agent and test specifications, the system (1) generates LLM agents, (2, 3) runs controlled web interactions, and (4, 5) analyzes logged behaviors for experiment owners.%
  }
  \Description{A flowchart diagram illustrates the A/B testing workflow enabled by UXAgent. On the left, a system user (e.g., UX researcher or product manager) provides two types of input: LLM Agent specifications (e.g., number of agents, demographic distributions, attributes) and A/B Testing configuration (e.g., feature designs, agent intentions, metrics). These inputs lead to two parallel processes: LLM Agent Generation (Step 1) and Testing Condition Preparation (Step 2). The generated agents are then assigned to either a control or treatment group (Step 3), where they interact with their designated web environments. The control group uses the baseline interface, while the treatment group uses a modified (new) design. Simulation results are collected (Step 4) and passed to the Post-Testing Analysis module (Step 5), which returns A/B Testing Results (e.g., session outcomes, average actions, failure rates) back to the user. The visual includes labeled arrows, icons representing agents and interfaces, and text boxes describing each component.}
  \label{fig:teaser}
\end{teaserfigure}

\maketitle

\section{Introduction}
\label{sec:introduction}

\textbf{A/B testing} (i.e., online controlled experimentation) is a cornerstone for evaluating design decisions in modern web applications~\cite{kohavi2009controlled,kohavi2007practical,kohavi2015online,goswami2015controlled}. 
By comparing alternative interfaces or features with users randomly assigned to control and treatment groups, it provides rigorous, data-driven evidence for user experience optimization~\cite{johari2017peeking,wang2019heavy,duan2021online}. 
Major technology companies, including Microsoft~\cite{machmouchi2017beyond}, Yahoo~\cite{tagami2014filling}, Netflix~\cite{amatriain2013beyond}, and Amazon~\cite{hill2017efficient}, run hundreds of such tests annually to refine features, improve engagement, and guide strategic design choices. 
Despite its ubiquity, traditional A/B testing is \revise{costly, slow to iterate, and difficult to revise when designs are flawed, due to long turnaround times and substantial engineering and organizational overhead~\cite{kohavi2013online,xu2015infrastructure,fabijan2018online}.}

Our goal is to better support A/B testing experimenters in designing and running experiments. To ground our system design, we conducted a formative study with six industry practitioners who regularly run large-scale A/B tests. Their accounts reveal three recurring bottlenecks (Fig.~\ref{fig:workflow}): (1) limited support for lightweight piloting before large-scale traffic allocation, (2) scarce and contested user traffic, and (3) slow feedback cycles between experiment setup and actionable insights. These constraints leave many promising ideas untested, and early-stage prototyping rarely benefits from rigorous, behaviorally grounded evaluation.

Recent advances in large language model (LLM) agents offer a promising way to address these limitations~\cite{chen2025towards,zhang2024revisiting,jia2024soul,zhang2025shop,chen2025unlearning,wang2025opera,wang2025customer,zhang2025see,wang2026trajectory2task}. 
LLM agents can role-play user personas and perform multi-step interactions in domains such as e-commerce~\cite{yao2022webshop}, booking~\cite{zhou2023webarena}, and information search~\cite{koh2024visualwebarena}. 
However, most prior work focuses on single-session tasks or constrained environments~\cite{li2024vital,luo2023wizardcoder,park2023generative,hua2023war,wu2023autogen,yao2023react}, limiting applicability to real-world websites. 
What remains missing is a scalable framework that deploys large numbers of persona-driven agents in live web environments to surface early behavioral signals before allocating real user traffic.

We address this gap with \textbf{\systemname} (Fig.~\ref{fig:teaser}), an end-to-end system for simulating large-scale online experiments with LLM-based agents. 
The system is interoperable and can plug into existing agent stacks (e.g., Claude computer-use agents or ReAct~\cite{yao2023react}). 
As shown in Fig.~\ref{fig:teaser}, \systemname integrates five components: agent specification, testing configuration, live agent–web interaction, behavioral monitoring, and automated analysis. 
At its core, LLM agents iteratively interpret dynamic webpages and ground persona-driven decisions into concrete DOM actions, enabling thousands of distributed sessions and scalable behavioral comparisons without manual intervention.

We emphasize that LLM-agent-based A/B testing is \textbf{not a replacement for real user testing}, but a complementary tool to help experiment owners (e.g., UX researchers and product managers) mitigate traffic scarcity, slow iteration cycles, and collaboration challenges~\cite{Kuang2022Merging, Feng2023CollabUX}.

We demonstrate \systemname through a case study on Amazon.com, simulating a between-subjects A/B test comparing the existing full filter panel with a reduced, similarity-based design. 
Using 1,000 agents (500 per condition), the reduced design produced more purchases than the baseline, with a modest but statistically significant increase. 
Spending showed a small upward trend, and agent outcomes aligned directionally with a parallel large-scale human A/B experiment, suggesting the system can surface meaningful behavioral signals.

In summary, this work contributes:
\begin{enumerate}
    \item \textbf{\systemname}, an end-to-end system for scalable, persona-driven A/B testing with LLM-based web agents.
    \item \textbf{Empirical evidence} from an Amazon.com case study showing alignment between agent simulations and human A/B results.
    \item \textbf{Design implications} for using agent-based simulation to support early prototyping and exploratory experimentation.
\end{enumerate}

\section{Related Work}
\label{sec:relatedwork}

We situate our work at the intersection of A/B testing, user behavior modeling, and HCI research on interactive prototyping and simulation.
We review two threads: (1) limitations of traditional A/B testing and advances in automated experimentation workflows, and (2) LLM-based agent simulation across domains.

\subsection{Limitations and Automation of A/B Testing}

A/B testing is a foundational methodology for comparing interface and feature variants to support data-driven design decisions~\cite{kohavi2009controlled,kohavi2007practical}, and is widely used in both industry and academia, including by companies such as Microsoft~\cite{machmouchi2017beyond}, Yahoo~\cite{tagami2014filling}, Netflix~\cite{amatriain2013beyond}, and Amazon~\cite{hill2017efficient}. 
Despite its ubiquity, prior HCI research documents persistent limitations, including slow iteration cycles, high development costs, limited early-stage feedback, and methodological challenges such as metric selection, data quality, and statistical reliability under real-world constraints~\cite{fabijan2017evolution,fabijan2018online,Gupta2019TopChallenges,Larsen2022StatisticalChallenges}. 
Beyond methodological issues, experimentation is shaped by socio-technical factors: UX practitioners face collaboration overhead, fragmented analyses, time pressure, and limited influence over data and model decisions in AI-intensive products~\cite{Kuang2022Merging,Feng2023CollabUX,Zdanowska2022MLUX,Lu2022AIUXTools}. 
In response, the HCI community has explored tools for rapid prototyping, in-situ sensemaking, and offline or automated experimentation~\cite{lasecki2015apparition,hartmann2006reflective,kuznetsov2022fuse,oppenlaender2020crowdui,gilotte2018offline,tamburrelli2014towards}, but these approaches still rely on historical data or live user traffic. 
Our work builds on these efforts by investigating agent-driven simulation as a complementary approach for evaluating interface designs without consuming real user traffic.

\subsection{User Behavior Simulation with LLM Agents}

HCI has long studied user behavior simulation through cognitive models such as GOMS and ACT-R~\cite{card2018psychology,olson1995growth,john1996using,gray2001cognitive}, and later through inverse reinforcement learning to infer user intentions from observed behavior~\cite{ziebart2009human,kangasraasio2018inverse}. 
While these approaches enable fine-grained modeling, they are labor-intensive, require domain expertise, and do not generalize well to open-ended tasks like online shopping; data-driven methods using server logs improve scalability but still rely on historical traces and fixed environments~\cite{paranjape2016improving}. 
Recent advances in large language models (LLMs) have renewed interest in behavior simulation across complex domains~\cite{ge2023openagi,wan2024building,wu2024sunnie}, with prior work demonstrating LLM agents’ ability to simulate social behavior~\cite{park2023generative}, reproduce behavioral experiments~\cite{horton2023large}, Agent-as-a-judge evaluation~\cite{chen2025multiagentasjudgealigningllmagentbasedautomated}, and improve action fidelity when trained on real-world data~\cite{lu2025can}. 
In parallel, web-agent benchmarks such as WebShop~\cite{yao2022webshop}, WebArena~\cite{zhou2023webarena}, VisualWebArena~\cite{koh2024visualwebarena}, WebVoyager~\cite{he2024webvoyager}, and WILBUR~\cite{lutz2024wilbur} have advanced autonomous web interaction, but remain largely confined to simulated or templated environments and focus on task completion rather than comparative user experience evaluation. 
Our work bridges these lines by deploying persona-driven LLM agents directly on live web interfaces to enable scalable user behavior simulation for A/B testing and early-stage design evaluation.

\section{Formative Study: Understanding Challenges in A/B Testing Workflows}
\label{sec:3-formative-study}

\subsection{Formative Study Method}

To inform the design of our LLM agent-based A/B testing system, we conducted a formative study with six industry professionals experienced in designing and running large-scale A/B tests. 
We used semi-structured interviews to elicit recent A/B testing experiences, focusing on workflows, tools, challenges, and how participants addressed them.
Participants were recruited via snowball sampling~\cite{biernacki1981snowball} from the e-commerce industry until thematic saturation. 
All six participants were based in the United States (four product managers, one software development manager, and one machine learning researcher) and routinely managed A/B tests with over one million users.
Interviews were conducted remotely, lasted 45–62 minutes, and were audio-recorded with consent. 
Recordings were transcribed, de-identified, and analyzed using grounded theory~\cite{muller2012grounded}. 
Two authors independently coded the data, reconciled differences, and synthesized themes related to A/B testing workflows, bottlenecks, and opportunities for system support.

\subsection{Formative Study Findings}
\label{subsubsec:3-formative-findings}

\paragraph{A/B Testing Project Lifecycle}

Interview data revealed a seven-stage A/B testing lifecycle (Fig.~\ref{fig:workflow}).
The process begins with \textbf{feature ideation}, where teams propose new UI changes or functionalities, followed by \textbf{team alignment and buy-in} involving discussion, refinement, and managerial approval.
Teams then conduct \textbf{experiment design}, specifying user segmentation, control and treatment conditions, and success metrics prior to development.
The \textbf{feature development and iteration} stage follows, requiring cross-functional collaboration and substantial engineering effort.
Once implemented, the \textbf{experiment launch} deploys the feature into a live A/B testing environment with randomized user exposure.
Afterward, \textbf{post-experiment analysis} interprets outcomes and behavioral drivers, leading to a final \textbf{feature decision} to launch or discard the feature.
Participants reported that this end-to-end process typically spans three months to over a year and requires close coordination across multiple roles.

\paragraph{Challenges and Coping Strategies.}
Interviewees highlighted recurring challenges in current A/B testing workflows, including the \textbf{high cost and long timelines of feature development}, which leave little opportunity for early user feedback before formal testing. To cope, some teams rely on internal alpha testing with colleagues, though participants noted this feedback is often biased and insufficient for guiding iteration, underscoring the need for lightweight prototyping and early-stage evaluation. Interviewees also described \textbf{intense competition for user traffic}, which forces experiments affecting similar interface components to be serialized and prioritized late in the process, delaying deployment and iteration. Finally, \textbf{high experiment failure rates} often result in one-shot decisions with limited opportunities for refinement, leading participants to call for tools that enable internal evaluation, data-driven prioritization, and predictive assessment of experiment outcomes prior to live A/B testing.

\section{\systemname}

\subsection{System Overview and Pipeline}

\systemname is an end-to-end system for simulating A/B tests on websites with persona-driven LLM agents. Experiment owners provide (1) agent specifications (e.g., number of agents and persona attributes), (2) A/B testing configurations (e.g., task intentions and behavioral metrics), and (3) two functional web variants for the control and treatment conditions (Fig.~\ref{fig:teaser}). Given these inputs, \systemname orchestrates four coordinated modules: (i) \textbf{LLM Agent Generation}, (ii) \textbf{Testing Preparation}, (iii) \textbf{Autonomous A/B Simulation}, and (iv) \textbf{Post-Testing Analysis}.

First, the \textbf{LLM Agent Generation Module} instantiates a population of agents by generating diverse personas and pairing them with the user-specified task intentions, ensuring variability in demographics and behavioral tendencies while adhering to the experiment constraints. The experiment owner specifies a target demographic distribution and provides an example persona as a style reference; a persona pool is initialized with that example. The system then iteratively samples an existing persona from the pool, samples demographic attributes from the target distribution, and prompts the LLM to produce a new persona matching both the sampled demographics and the reference style. Each generated persona is added back to the persona pool, enabling diverse yet stylistically consistent personas at scale. Second, the \textbf{Testing Preparation Module} assigns agents to control and treatment conditions via traffic allocation and checks balance across key persona attributes, re-splitting when needed to reduce distributional skew. Third, in the \textbf{Auto\-nomous A/B Simulation Module}, agents interact with their assigned web variant in isolated browser sessions using a recurrent perceive--decide--act loop (Section~\ref{sec:agent}), while the system records step-level traces of agent actions and web states. Finally, the \textbf{Post-Testing Analysis Module} aggregates traces across conditions to compute outcome metrics (e.g., completion or purchase rates), interaction efficiency measures (e.g., actions per session, session duration), and behavior patterns (e.g., search or filter usage), and supports stratified analyses by persona attributes to surface subgroup differences. Together, these modules enable scalable comparative evaluation of interface variants and provide early behavioral signals without relying on live user traffic.
Full details of the system design can be found in Appendix \ref{sec:system}.

\begin{table*}[t]
    \begin{booktabs}{colspec={lccc}}
    \toprule
    \textbf{} & {\textbf{Control Condition} \\ Full Filter List \\ \textbf{Human}, N=1M} & {\textbf{Control Condition} \\ Full Filter List \\ \textbf{Agent}, N=500} & {\textbf{Treatment Condition} \\ Reduced Filter List \\ \textbf{Agent}, N=500} \\ 
    \midrule
    
    \textbf{Search}               & 6.40  & 1.42  & 1.43            \\ 
    \textbf{Click\_product}          & 6.96  & 1.87  & 2.09             \\ 
    \textbf{Click\_filter\_option}    & 0.33  & 0.58  & 0.60             \\ 
    \textbf{Purchase}             & 0.62  & 0.81  & 0.83             \\ 
    \textbf{Stop}                 & -      & 0.19  & 0.17             \\ 
    \midrule
    \textbf{Average actions}          & 15.96 & 6.05 & 6.60              \\ 
    \textbf{\# of purchase *} & - & \textbf{403} & \textbf{414} \\
    \textbf{Average \$ spend} & - & \$55.14 & \$60.99 \\
    \bottomrule
    \end{booktabs}
    \caption{Comparison of human customers' actions per session in control condition (current design with full filter list), and the LLM agents as virtual customers in both control condition and treatment condition (new design with reduced filter list). 
    The LLM agents in treatment condition purchased significantly more items than the ones in the control condition.  $\chi^2(1) = 5.51$, p-value < 0.05
    }
    \vspace{-2\baselineskip}
    \label{tab:agent-vs-human}
\end{table*}

\subsection{Agent-Environment Interaction Architecture}
\label{sec:agent}

At the core of \systemname is an iterative interaction loop in which each LLM agent operates directly on a live web environment and continuously adapts its actions based on evolving page state. This loop consists of three tightly coupled components: an \textbf{Environment Parsing Module}, an \textbf{LLM Agent}, and an \textbf{Action Execution Module} (Fig.~\ref{fig:agent}). Together, they enable robust and scalable interaction with dynamic, real-world webpages.

At each step, the \textbf{Environment Parsing Module} extracts the current webpage into a structured representation of task-relevant elements (e.g., products, filters, prices) and the available action space, reducing noise from raw HTML or visual clutter. This observation, together with the agent’s persona, intention, and interaction history, is passed to the \textbf{LLM Agent}, which performs language-based reasoning and planning to select the next structured action (e.g., search, click filter, select item, purchase), independent of any specific agent architecture. The \textbf{Action Execution Module} then translates this action into browser-level interactions and executes it in the live environment, incorporating basic fault handling such as retries or re-parsing to accommodate dynamic page changes. This perceive–decide–act loop continues until task completion or failure, while \systemname logs fine-grained traces of page states, actions, and outcomes for downstream A/B-style analysis.

\section{Case Study: A/B Testing with Reduced Filter Options on Amazon.com }
\label{sec:experiment}

We evaluate \systemname through a simulated A/B testing case study on Amazon’s shopping interface, focusing on a redesign of the left-side filtering panel, a common target for UI experimentation in e-commerce. The treatment condition replaces the full filter list with a similarity-based ranking design that suppresses low-relevance options, while the control condition retains the existing interface. Using \systemname, we generated a large pool of agent personas and sampled 1,000 agents (500 per condition) to simulate shopping sessions under realistic demographic distributions and task intentions, in parallel with a large-scale human A/B test on the same feature. Agents interacted with the live website using standard e-commerce actions, and the system logged full interaction traces and outcomes for analysis. Full details of the scenario design, agent setup, infrastructure, and cost analysis are provided in Appendix~\ref{sec:scenario}.

\subsection{Finding: Alignment with Human Behavior}

To assess how well \systemname aligns with real user behavior, we compared a simulated A/B test using 1,000 LLM agents with results from a parallel online A/B experiment involving 2M human participants under identical task conditions (Table~\ref{tab:agent-vs-human}). Humans exhibited more exploratory interaction patterns, including longer sessions and more frequent searches and product views, while agents followed more goal-directed trajectories with fewer actions. Despite these differences in interaction style, agents and humans showed comparable purchase rates and similar use of filter options, indicating alignment in decision outcomes for intention-driven tasks. Overall, these results suggest that while LLM agents simplify exploratory behavior, they capture key decision-making signals relevant for early-stage UX evaluation, supporting the use of agent-based simulation as a complementary tool to human A/B testing.

\subsection{Findings: System Effectiveness Across Interface Variants}

To evaluate whether \systemname can detect differences between interface variants, we compared LLM agent behavior across control and treatment conditions with different filter designs. As shown in Fig.~\ref{fig:all-bars} and Table~\ref{tab:agent-vs-human}, agents in the treatment condition interacted more with products and filters and completed more purchases than those in the control condition (414 vs.\ 404; $\chi^2(1)=5.51$, $p=0.03$), indicating a modest but statistically reliable increase. Average spending exhibited a small upward trend under the treatment design, though this effect did not reach statistical significance.

Subgroup analyses further revealed heterogeneous responses across personas (Fig.~\ref{fig:all-bars}): agents representing older and male customers showed larger increases in spending under the simplified filter design, while younger agents exhibited decreased spending, suggesting differing preferences for filter complexity. Importantly, these agent-based patterns aligned directionally with results from human A/B testing, supporting \systemname’s ability to surface meaningful behavioral signals and detect subtle interface effects prior to live deployment.

\section{Discussion}
\label{sec:discussion}

Our design and evaluation of \systemname demonstrate how agent-based simulation can serve as a design support mechanism in the web interface design lifecycle. Rather than replacing real-user experimentation, \systemname complements existing workflows by enabling earlier, faster, and lower-risk exploration, aligning with broader trends in HCI that leverage large-scale simulation for iterative design and decision-making~\cite{park2023generative,horton2023large,gao2024large}.

\paragraph{Simulation-Supported Design Iteration and Exploration.}
A key contribution of \systemname is its ability to accelerate design iteration by enabling agent-based pilot experiments prior to A/B testing. Although A/B testing remains the gold standard for design validation, it is constrained by limited user traffic, long feedback cycles, and substantial engineering and organizational overhead~\cite{fabijan2017evolution,kohavi2013online,xu2015infrastructure}. By supporting rapid evaluation of interface variants through agent simulations, \systemname allows designers to obtain early signals without implementing and deploying their new design to real users. This capability extends prior prototyping and experimentation tools~\cite{lasecki2015apparition,hartmann2006reflective,kuznetsov2022fuse} by grounding early feedback in end-to-end interaction traces, and enables simulation-driven methodologies such as parallel evaluation and computational design exploration~\cite{kumar2013webzeitgeist}.

\paragraph{Inclusive Piloting and Risk-Aware Testing.}
Beyond accelerating iteration, \systemname supports inclusive and risk-aware piloting for user populations that are difficult, costly, or ethically sensitive to recruit. Groups such as older adults or users with lower digital literacy are often underrepresented in early experimentation, even though design flaws may disproportionately affect them. By configuring agent personas to reflect diverse demographic and behavioral characteristics, designers can explore how features may impact different user groups before exposing real users to potential harm. This capability complements prior crowd-based and proxy evaluation approaches~\cite{bernstein2011crowds,oppenlaender2020crowdui} by offering an automated alternative that helps surface subgroup disparities early, enabling more robust and informed A/B testing, and aligns with emerging evaluation techniques using agent judges~\cite{chen2025multiagentasjudgealigningllmagentbasedautomated}.

\section{Conclusion}

In this paper, we presented \systemname, a system that enables large-scale, LLM agent-based simulation of A/B testing for web interfaces. 
Our evaluation demonstrated that LLM agents exhibit realistic, goal-aligned behaviors, and are sensitive to interface variations.
By supporting rapid, risk-free behavioral piloting, \systemname introduces a new phase for agent-based piloting in the design life-cycle that complements traditional A/B testing and expands the scope of early-stage UX evaluation. 
We envision future extensions that further enhance agent fidelity, broaden domain coverage, and integrate simulation into intelligent design optimization workflows.

\bibliographystyle{ACM-Reference-Format}
\bibliography{bib/sample-base,bib/software,bib/custom,bib/yuxuan}

\appendix

\section{\systemname}
\label{sec:system}

\begin{figure*}[t]
    \centering
    \includegraphics[width=.95\linewidth]{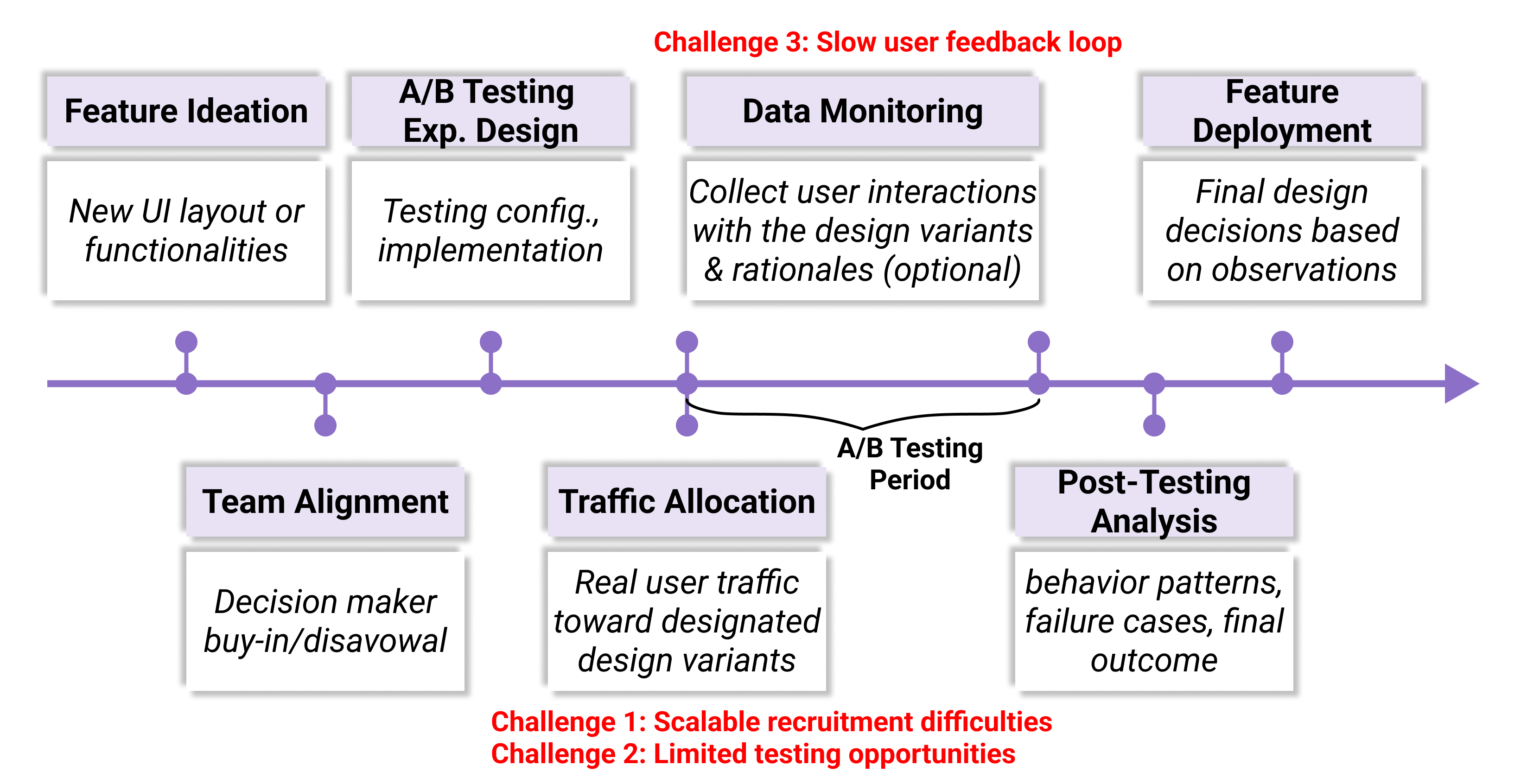}
    \caption{The workflow of web A/B testing and three challenges reported from formative study: (1) the cost and difficulty of securing considerable user traffic for significant results, (2) the whole A/B testing period can span across weeks and months,  and (3) limited testing opportunites.    }
    \Description{A timeline diagram illustrates the standard A/B testing workflow and highlights three key challenges. The timeline includes seven sequential phases: Feature Ideation (new UI layout or functionalities), A/B Testing Experiment Design (testing configuration and implementation), Data Monitoring (collect user interactions and rationales), Feature Deployment (final design decisions), Post-Testing Analysis (behavior patterns, failure cases, outcomes), Traffic Allocation (real user traffic toward variants), and Team Alignment (decision maker approval or rejection). The A/B Testing Period is marked across the Data Monitoring and Traffic Allocation stages. Three challenges are highlighted in red: (1) Scalable recruitment difficulties, (2) Limited testing opportunities, and (3) Slow user feedback loop. Each stage is connected along a central purple timeline.}
    \vspace{-1\baselineskip}
    \label{fig:workflow}
\end{figure*}

\begin{figure*}[t]
    \centering
    \includegraphics[width=\textwidth]{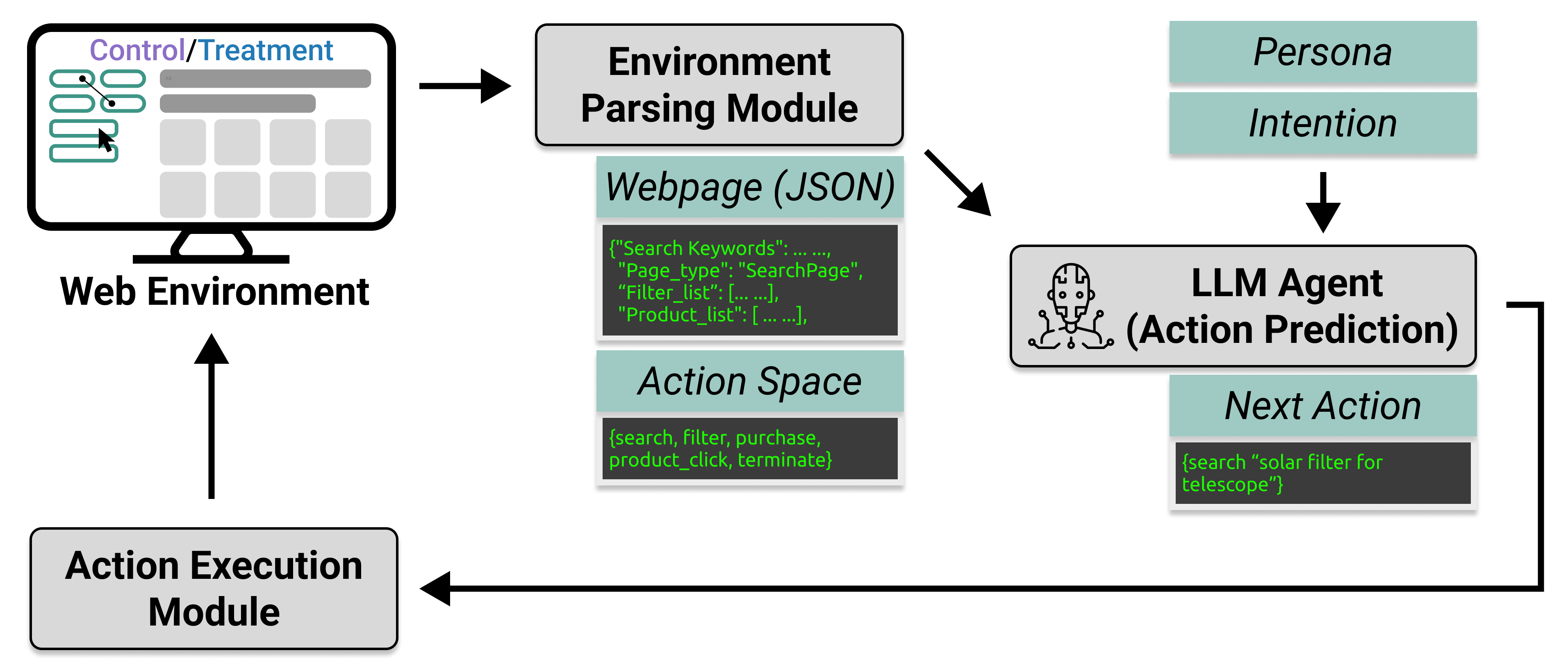}
    \caption{One action prediction iteration of the automated web testing in \systemname with an LLM agent. 
    (1) An \textit{Agent Profiling Module} maintains a comprehensive agent description with an LLM-generated persona, user-specified intention, and the action history of the current session. In the meantime, (2) the \textit{Environment Parsing Module} parses the webpage into structured web representation and action spaces. (3) All the information is fed into the \textit{LLM Agent} for the next action prediction, (4) which will be executed by the \textit{Action Execution Module} in the web environment to drive the next iteration. \revise{The light-weight Environment Parsing Module and Action Execution Module are web-site specific by their nature, and other modules include LLM Agent are generalizable to different websites.}
    }
    \Description{
A system diagram shows how a language model-based agent interacts with a web environment. The process begins with a web interface (Control/Treatment) feeding into an Environment Parsing Module, which extracts the webpage as a JSON structure and defines an action space (e.g., {search, filter, purchase, product_click, terminate}). This information, along with a predefined persona and intention, is sent to the LLM Agent (Action Prediction), which outputs the next action (e.g., {search “solar filter for telescope”}). This action is executed through the Action Execution Module, which sends it back to the Web Environment, completing the interaction loop.}
    \label{fig:agent}
\end{figure*}

\subsection{System Overview and Pipeline}

\systemname is designed as an end-to-end simulation system for LLM agent-based A/B testing in live browser environments. 
Analogous to the preparation stage of real user A/B testing, the \revise{experiment owners} (i.e., UX researchers or product managers) need to determine details of A/B testing designs and provide two web environment variants to be tested.
After taking the user inputs, the system operates on four LLM-powered modules (Fig.~\ref{fig:teaser}): LLM agent generation, testing preparation, autonomous A/B testing simulation, and post-testing analysis. 

\systemname system users begin by specifying the \textbf{LLM agent specifications} and \textbf{A/B testing configurations}. The agent specification defines the target user population, including the number of agents, demographic and behavioral diversity (e.g., age, education, tech literacy), and other persona attributes. These personas drive the agents' planning and reasoning processes and introduce behavioral variability across the simulation.

In parallel, users configure the A/B test by defining initial user intentions, behavioral metrics to track, and the design features to evaluate. User intentions guide the agent's interaction trajectory and termination condition (e.g., ``find a discounted Bluetooth speaker under \$30''). The design features (such as layout or interaction flow) are implemented in fully functional web environments. In our e-commerce scenario, the tested design feature is the layout of the filter panel on the website.

Once these inputs are provided by the experiment owners, the \textbf{LLM Agent Generation Module} queries the backend LLM to generate the specified number of LLM agents with diverse persona descriptions and intentions.
The query explicitly instructs the LLM to ensure the generated LLM agent persona and intentions must follow the user-provided agent specifications.
In our experiment in Section~\ref{sec:experiment}, we leveraged this module to generate \textbf{100,000 agents}.
We provide a sample persona generated by our system in Appendix~\ref{app:example}.
After generating the LLM agents as the A/B testing participants, the \textbf{Testing Preparation Module} performs the agent traffic allocation by splitting the agents into control (without new feature) and treatment (with new feature) groups, and each group is assigned to interact with the corresponding web environments. 
The statistics of agent characteristic distributions will be calculated within each group to ensure that the distribution of the LLM agents is balanced; if the statistics are not balanced, the Testing Preparation Module will re-execute agent traffic allocation until satisfactory.

These web environments for both groups need to be launched using independent browser instances controlled through ChromeDriver (for web environment parsing) and Selenium WebDriver (for automated interaction) integration.
During the \textbf{automatic interaction with the web environment}, each LLM agent begins to interact with the webpage using an autonomous action prediction loop, which is shown in Fig.~\ref{fig:agent}.
Each loop involves perceiving the webpage state, interpreting the action space, predicting the next action, and executing that action in the browser. .
During the process, each step of the interaction is recorded, and the system monitors the overall session progression until termination, such that the agent accomplishes the intended goal or encounters failure cases.

After all agents complete their interactions with the assigned web environments, \systemname transitions to the result analysis stage. 
The \textbf{Post-Testing Analysis Module} is responsible for aggregating, interpreting, and presenting agent behaviors in a form that supports A/B-style experimental comparison. 
The output of this module serves as the primary feedback surface for the system user.
Each agent session produces a fine-grained action trace that includes the full sequence of interactions, timestamps, webpage states, executed actions, intermediate rationales (when available), and final outcomes. These logs are collected asynchronously during the simulation and stored in a structured format. Upon session termination, the analysis module aggregates these records across both the control and treatment groups to extract comparative metrics and visualize key behavioral dynamics.

The\textbf{ post-test analysis} module outputs summary statistics such as actions per session, session duration (in steps and time), and purchase completion rate. Researchers can also examine detailed behaviors (e.g., search or click filter usage) and compare them across A/B condition variants. The system supports stratified analysis by agent demographics or personas to identify subgroup differences. 
For instance, when testing redesigned filters, the system can reveal whether agents refined searches more, completed tasks faster, and purchased more, which offers early insights on usability and adoption risks before live deployment.

Finally, \systemname maintains compatibility with external analytics pipelines. The result logs (in .JSON and .XLS) are exported in a format compatible with common data science tools, enabling downstream statistical modeling, significance testing, or integration with traditional A/B testing dashboards. Users can use \systemname to analyze thousands of simulated interaction session results into actionable insights for interface evaluation.

This pipeline allows \systemname to support fully automated LLM-agent-web interactions at scale under various environment configurations. 
A single experimental run can involve hundreds or thousands of sessions distributed across different personas and design conditions, all executed without any human intervention.

\subsection{Agent-Environment Interaction Architecture}

At the core of \systemname is an iterative mechanism where each LLM agent continuously interacts with a real web environment by dynamically updating its understanding of the environment and adjusting its actions accordingly.
This architecture consists of three tightly integrated components: the Environment Parsing Module, the LLM agent, and the Action Execution Module. 
Together, these components enable robust operation in complex and dynamic web environments, for instance, our system evaluation in Section~\ref{sec:experiment} is experimented on the Amazon\footnote{\url{http://www.Amazon.com}} platform. \revise{We will release our code upon acceptance of this paper.}

\paragraph{Environment Parsing Module.}
The interaction begins in a dedicated browser-based web environment session.
Traditional approaches explored the extraction of webpages into screenshots or raw HTML representations.
However, these approaches are impeded by extracting overly complicated information with a lot of unwanted raw webpage information. For instance, the raw HTML and DOM trees retain complicated hierarchical information of elements in a webpage, and the screenshots introduce irrelevant visual content and increase processing latency.
In our work, the Environment Parsing Module in our system parses the web environment into structured observations with a JSON format that simplifies the structure of the website and stores only key information for the agent-web interactions.
In particular, we use a ChromeDriver to execute a JavaScript processing script within the browser. 
This script selectively extracts targeted information directly from the raw HTML by extracting essential web elements.

For the e-commerce scenario, we specifically designed the script to extract web elements like product filters, titles, descriptions, and customer ratings based on their unique identifiers (IDs or classes). 
On the search result page, as shown in Fig.~\ref{fig:design-filters}, for instance, we extract product details such as titles, names, ratings, reviews, and prices from the results section and gather filter options (e.g., Brand, Price, Delivery Day) from the left search panel.  
This approach is designed to remove irrelevant elements that are not relevant to the targeted design features and user interactions, such as advertisements, banners, or unrelated hyperlinks.
The JSON file generated by the Environment Parsing Module provides a cleaner and more focused observation of the web environment to the LLM agent.

The Environment Parsing Module also identifies the current action space, which defines the set of allowable actions an agent can perform in the given context. 
These actions mimic the sequential steps a user would take while interacting with the website, such as typing keywords and clicking on items. 
Each action is structured in a format that the LLM agent can interpret and execute as part of its decision-making process.

The actions we define are represented in text format, which allows the agent to respond and perform tasks consistently. 
The key actions included are: (1) Search: The agent uses the search bar to find specific items or information. (2) Click Product: Select an item from the current webpage to view more detailed information about it. 
(3) Click Filter Option: Apply one of the available filters (e.g., price range, brand) to refine the search results. 
(4) Purchase: Complete the purchase of the selected item. 
(5) Stop: Indicate that the shopping session is complete and no further actions are required. This approach supports realistic simulation of shopping processes, providing a basis for evaluating LLM capabilities and examining human-like behavior in online shopping environments.

\paragraph{LLM Agent.}
The structured webpage and action space are passed to the LLM agent, which also receives the agent's persona and intention--the initial intention was generated by the LLM Agent Generation Module but could be dynamically updated within the LLM Agent. 
The intention specifies the agent's current task (e.g., searching for a specific product, comparing alternatives, or making a budget-constrained purchase). 

The LLM agent functions as a decision-making module that consumes the current state and outputs the next action to be taken. 
In particular, the LLM agent models the next-step decision-making problem as a form of language-based reasoning and planning task by mapping structured state observations into reasoning traces and action predictions. 
\systemname is not bound to a specific LLM agent. 
Instead, our system treats the LLM agent as an exchangeable module that supports various types of LLM web agents (ReAct \cite{yao2023react}, FireClaw\footnote{\url{https://github.com/mendableai/firecrawl-mcp-server}}) with convenient ``plug-and-play'' APIs, analogous to the Model Context Protocol (MCP) proposed by Claude.

\begin{figure}[t]
    \centering
    \includegraphics[width=\linewidth]{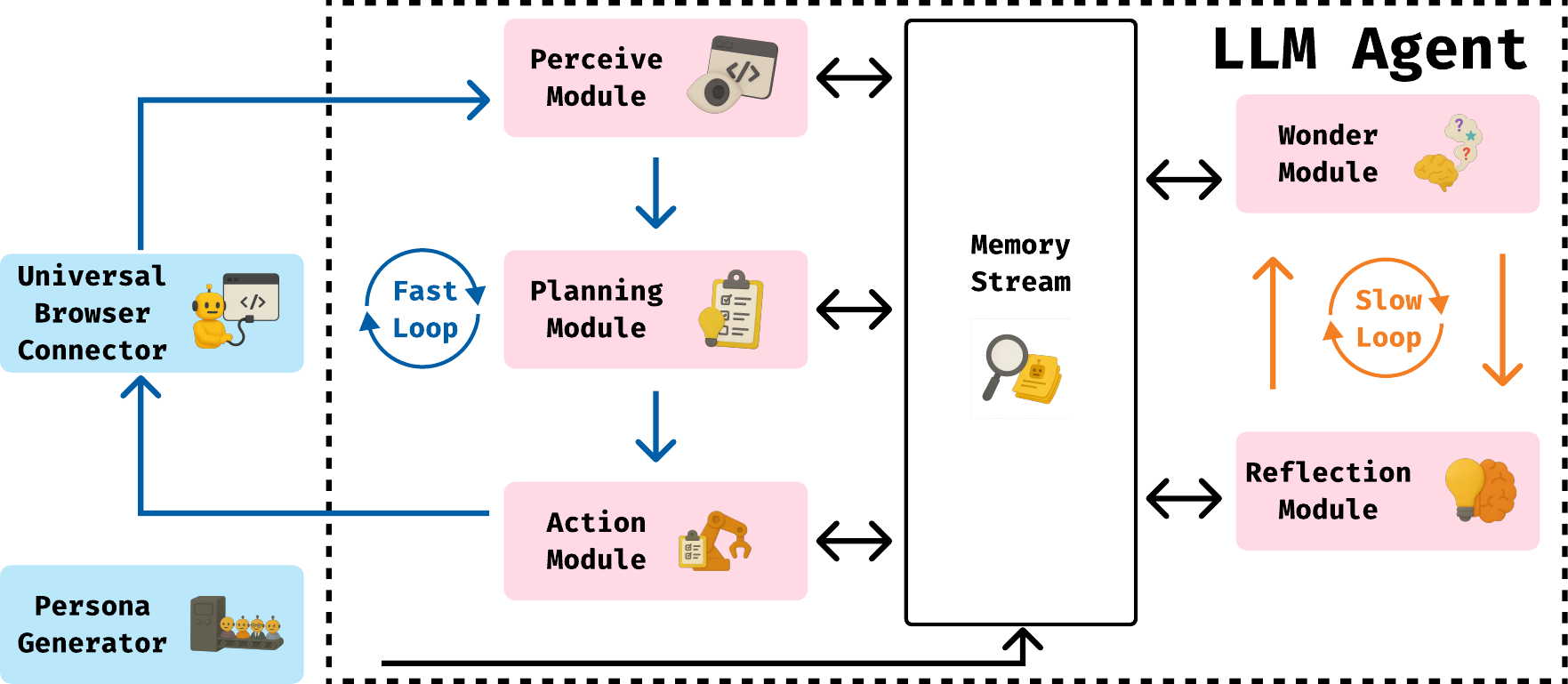}
    \caption{\revise{UXAgent Architecture design \cite{luUXAgentLLMAgentBased2025a}.}
    }
    \label{fig:structure}
    \vspace{-1\baselineskip}
    \Description{The architecture features two interacting processes: the Fast Loop and the Slow Loop, which handle rapid web interactions and deeper reasoning respectively. Within the fast loop, the perception module converts raw observations from the environment into natural language memory entries stored in the memory stream, while the planning module generates short-term plans that the action module executes.
    In the slow loop, the wonder module simulates human ``mind wandering’’ and the reflection module performs more thorough reasoning about the current context.
    Modules retrieve relevant entries from the memory stream and also append newly generated memories back into it.}
\end{figure}

\revise{In our experiment, we adopt the UXAgent framework~\cite{luUXAgentLLMAgentBased2025a}, a state-of-the-art LLM agent for web interactions, as a representative implementation due to its strong performance and its built-in support for multi-step planning, hierarchical control, and intermediate memory management. UXAgent (Fig. \ref{fig:structure}) follows a dual-loop architecture composed of a Fast Loop for rapid perception–action cycles and a Slow Loop for deeper reasoning. Within the fast loop, the perception module converts raw signals from the browser connector into natural-language observations, which are then appended to the memory stream. The planning module generates short-horizon plans conditioned on these observations, and the action module executes the action. The slow loop complements this by incorporating a wonder module that generates speculative or exploratory thoughts, simulating human-like mind wandering, and a reflection module that performs deliberate reasoning over past trajectories and current goals. Both loops continuously read from and write to the unified memory stream, allowing perception, planning, action, wonder, and reflection to jointly shape the agent’s evolving internal state and decision process.}

\paragraph{Action Execution Module.}
The next action predicted by the LLM agent is translated into browser commands by the Action Execution Module. 
Actions are expressed in a structured format that can reference DOM elements or logical operations, such as \textbf{Click\_product(3)}, \textbf{Click\_filter\_option(Brand: Sony)}, \textbf{Search("Wireless earbuds")}, or \textbf{Purchase}. 
The execution module parses the action and performs the corresponding interaction on the live webpage. 
In some cases, the web execution is not guaranteed to succeed, as real-world pages are prone to dynamic content loading and modal interruptions. 
Therefore, the execution module incorporates built-in fault detection and recovery logic. 
If an action fails due to a missing selector or DOM mismatch, the system attempts fallback options such as retrying, scrolling into view, or re-parsing the page. 
Each execution updates the environment and starts the next iteration of the loop.

This interaction loop continues until the agent reaches a terminal condition. 
Successful termination occurs when the agent completes its task—for example, by navigating to a purchase page or explicitly declaring task success. Failure conditions include endless loops, unreachable goals, or repeated in-executable actions. Sessions are also capped with time and action count thresholds to prevent infinite rollouts. 
Each completed session produces a full trace of interaction history, action rationales, page states, and final outcomes.

\section{A/B Testing Scenario}
\label{sec:scenario}

\begin{figure*}[t]
    \centering
    \includegraphics[width=.95\textwidth]{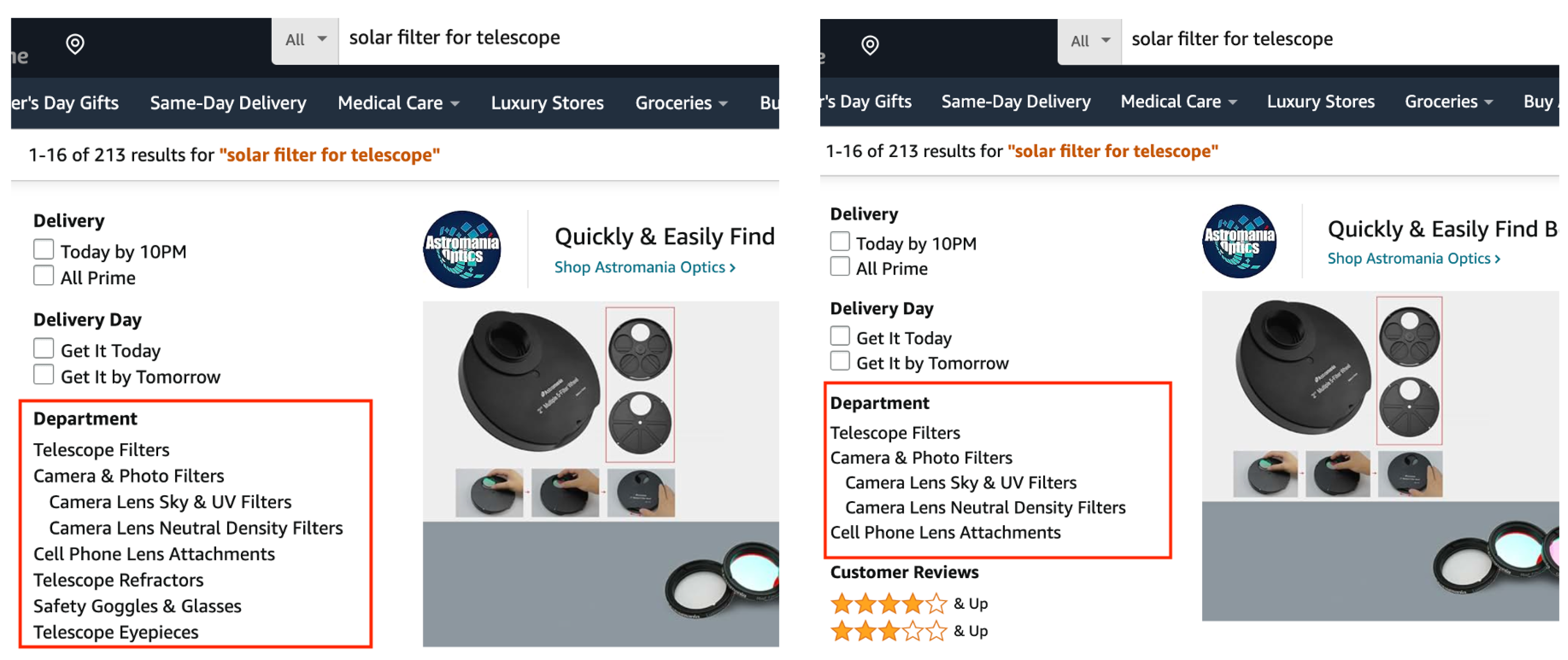}
    \caption{The two design variants of the left filter panel on Amazon.com for the A/B testing case study with \systemname. 
    In the control condition (left screenshot), all the filter options are shown; in the treatment condition (right screenshot), we apply a similarity-based ranking algorithm to reduce the filter options that have lower than 80\% similarity to the user's search query (e.g., ``solar filter for telescope'').}
    \label{fig:design-filters}
    \Description{A side-by-side comparison of two Amazon search results pages for the query "solar filter for telescope." Both pages show similar products but have differences in their filter options on the left sidebar. The left image displays additional filter categories under "Department" such as "Telescope Refractors," "Safety Goggles \& Glasses," and "Telescope Eyepieces," which are not present in the right image. Both images include nested subcategories under "Camera \& Photo Filters" and identical product listings on the right side of the page. This comparison highlights subtle differences in the user interface that could affect user navigation or filtering behavior.}
\end{figure*}

We designed a simulated A/B testing scenario focused on a novel feature design of the left-side filtering panel of Amazon’s shopping interface, a common area for UI experimentation in e-commerce. 
Our team has the privilege of altering the left filter panel interface (Fig.~\ref{fig:design-filters}). In comparison to the existing design as the control condition, where all the filter options are shown, the new design (treatment condition) uses a similarity-based ranking algorithm to reduce the filter options that have a lower-than-80\% similarity score to the user's search query.

\revise{Using the LLM Agent Generation Module, we created 100{,}000 agent personas\footnote{We will open source these agent personas} and then randomly sampled 1{,}000 of them to simulate individual shopping sessions, with 500 assigned to the treatment condition and 500 to the control condition. The personas' demographic attributes were sampled to match real online shopper distributions, including age (18–24: 15.8\%, 25–34: 24.6\%, 35–44: 22.1\%, 45–54: 19.6\%, 55–64: 17.9\%), gender (49.4\% male, 50.6\% female), and household income (0–30{,}000: 32.1\%, 30{,}001–94{,}000: 32.4\%, 94{,}001–1{,}000{,}000: 35.5\%).}
In parallel, we also conducted the real human A/B testing with 2 million users randomly assigned into the control condition and the treatment condition.
Each agent was initialized with a persona profile and a shopping task (e.g., ``find a budget smart speaker under \$40 with strong customer reviews''). The allowed action space for the LLM agent includes typical e-commerce flows: search, filter use, product examination, and cart actions. Persona generation followed the methodology in \citet{chen2024empathy} with LLM agent's demographics (age, income, occupation), preferences, and shopping goals.

In the case study, our \systemname was implemented with the Claude 3.5 Sonnet model as the LLM backend to support agent generation, testing preparation, automated agent-web interaction, and post-testing analysis.
The \systemname environment was executed on a distributed cluster of 16 high-memory compute nodes, with each node controlling a Selenium-driven Chrome instance running in headless mode. 
Each session was capped at 20 actions. 
Sessions ended either with task success (completion or no further predicted actions) or failure (e.g., looping behavior). We logged full action traces, metadata (e.g., duration, outcome), and the agent's rationales where applicable.

\revise{To learn the environment cost of LLM Agents (as suggested by \cite{10.1145/3706598.3714227}), we calculated the number of tokens and CO2 emissions of our experiment. For 1,000 agent simulations using Claude 3.5 Sonnet, estimated 875 million tokens, resulting in a cost of approximately \$2,925 and CO2 emissions of 35 kilograms (using 0.00004 g CO2/token for average data center grid mix). In comparison, recruiting 1,000 human participants into a typical UX study session in the AI Industry would cost \$100 per person (\$100,000 in total).}

\begin{figure}
\centering
\includegraphics[width=\linewidth]{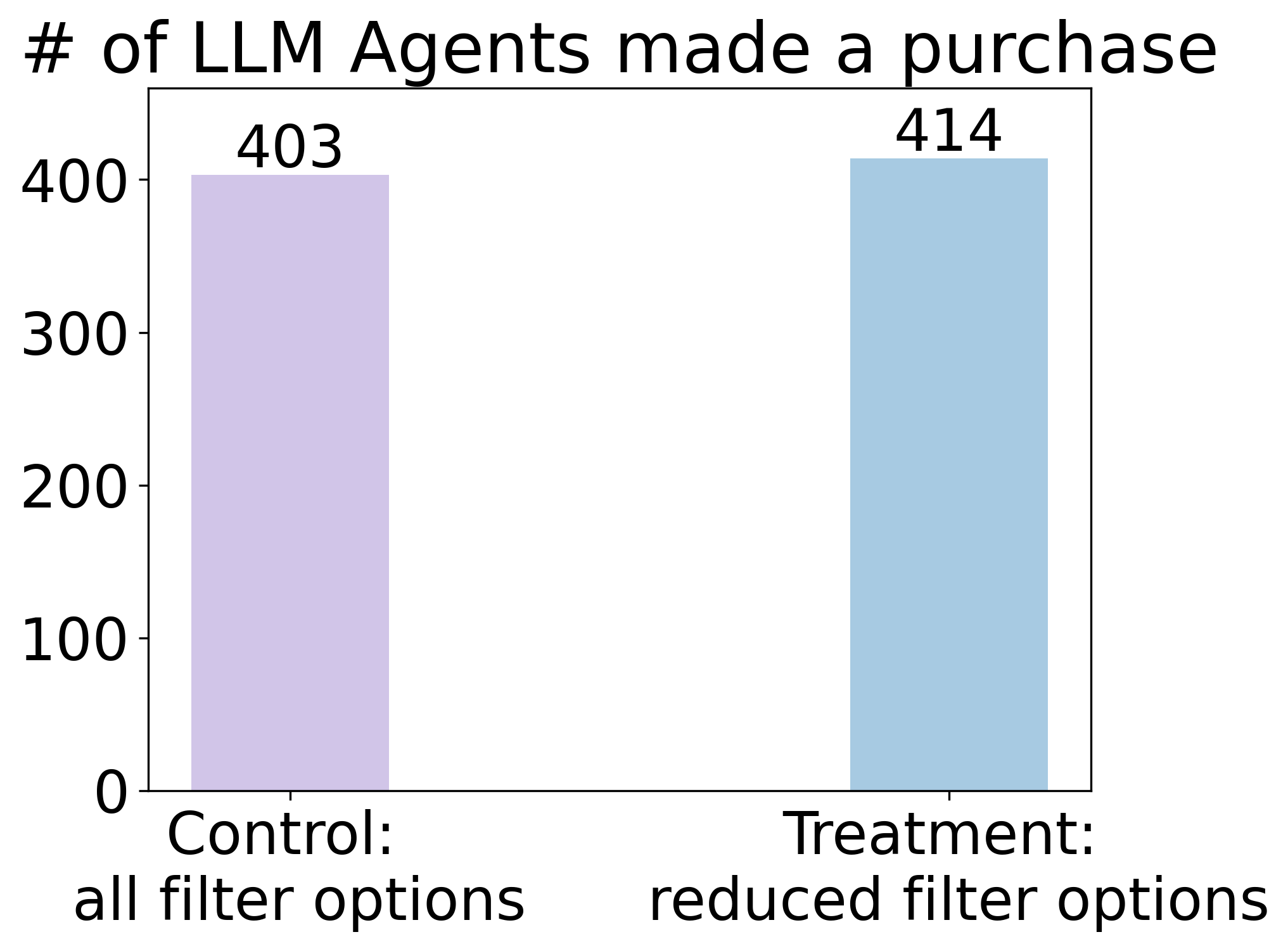}
\caption{Number of LLM agents who completed a purchase under control and treatment conditions.} 
\Description{A bar chart comparing the number of LLM agents who made a purchase. The control group with all filter options has 403 purchases, while the treatment group with reduced filter options has 414 purchases.}
\label{fig:placeholder}
\vspace{-1\baselineskip}
\end{figure}

\begin{figure*}[t]
    \centering
    \begin{subfigure}{0.45\linewidth}
        \centering
        \includegraphics[width=\linewidth]{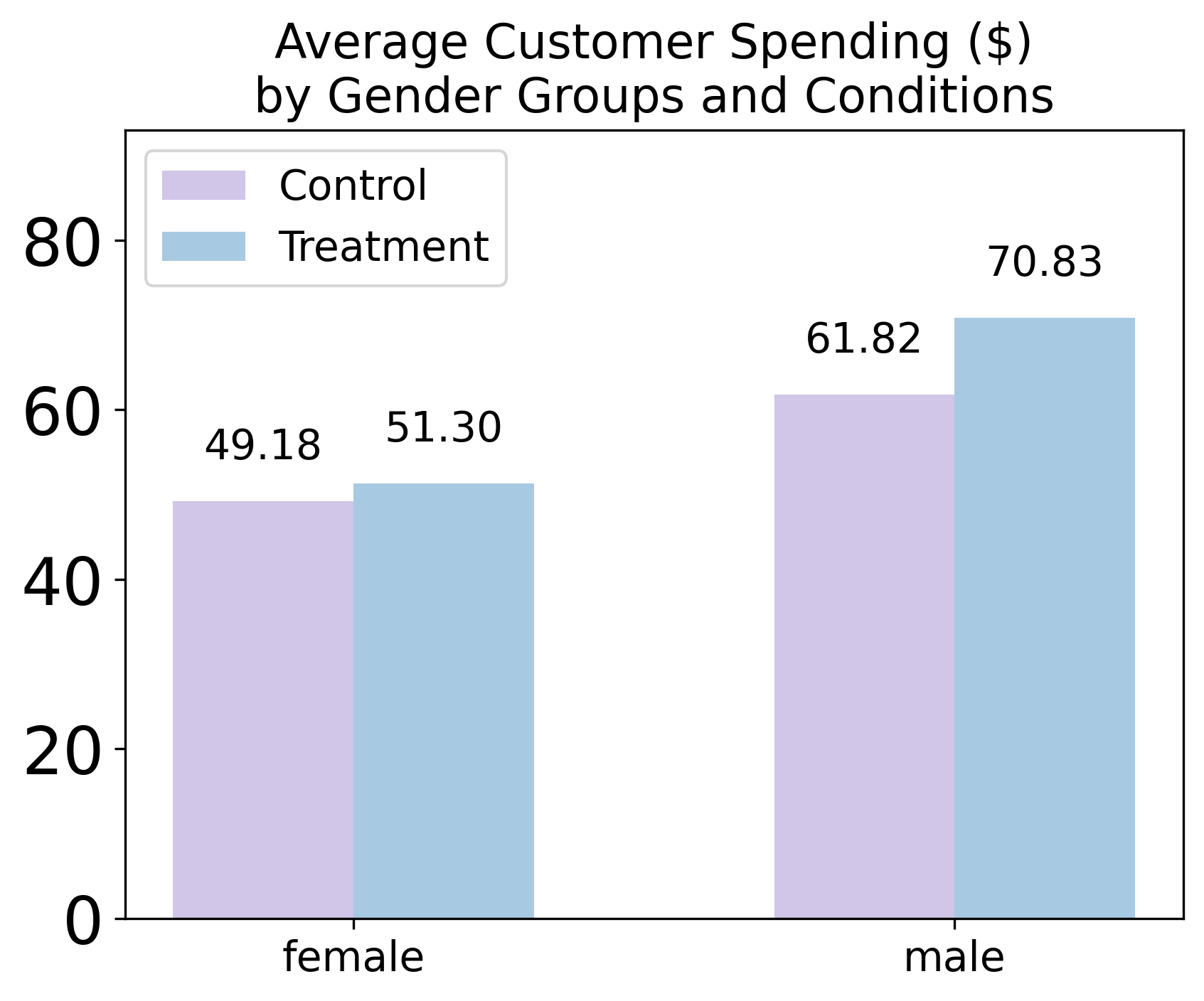}
        \caption{Average customer spending by gender groups across conditions.}
        \label{fig:control-treatment-gender}
    \end{subfigure}
    \hfill
    \begin{subfigure}{0.45\linewidth}
        \centering
        \includegraphics[width=\linewidth]{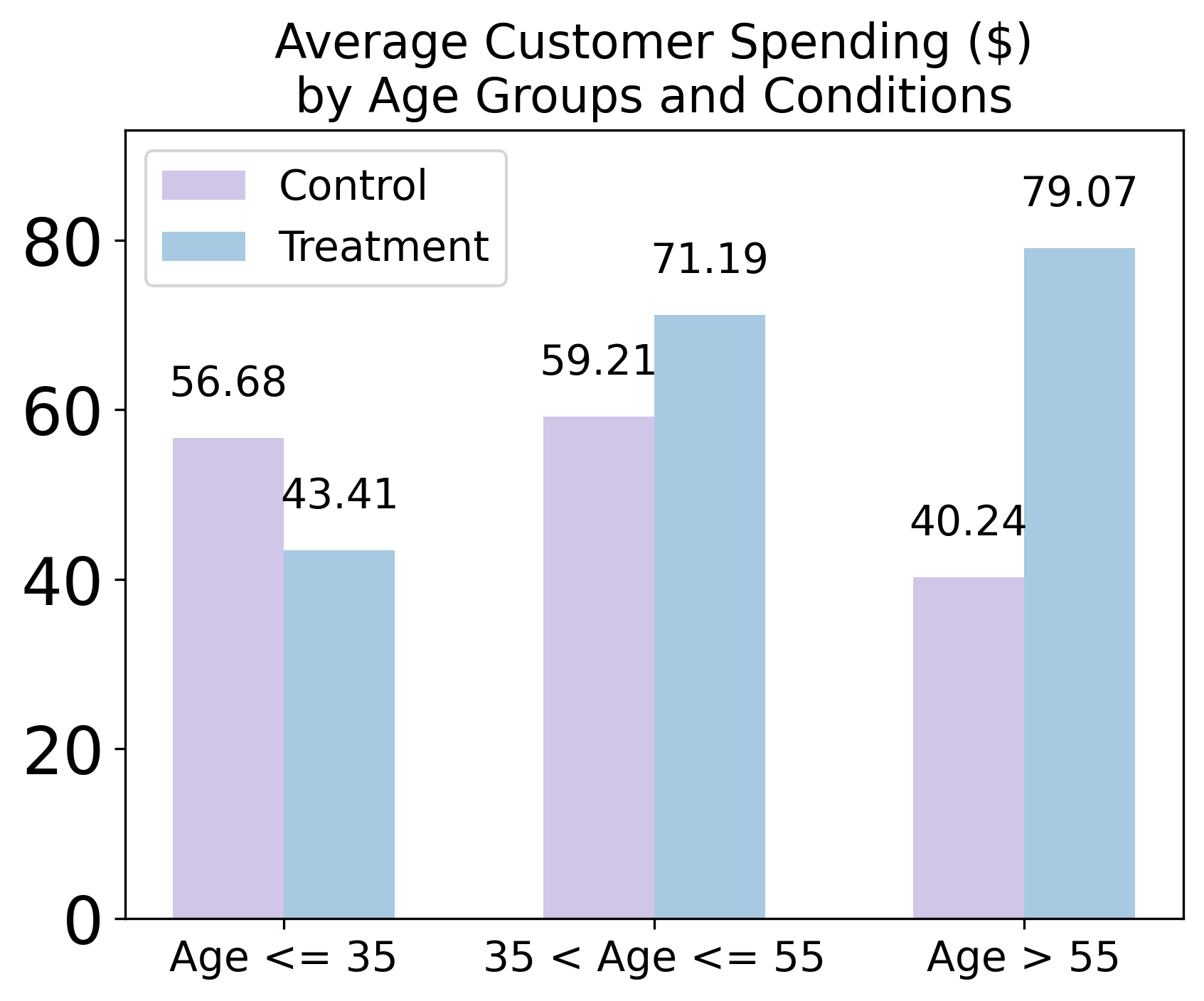}
        \caption{Average customer spending by age groups across conditions.}
        \label{fig:control-treatment-age}
    \end{subfigure}
    \caption{Comparison of average customer spending across conditions, broken down by gender (a) and by age groups (b).}
    \Description{Two bar charts compare average customer spending in U.S. dollars between control and treatment groups.

Chart (a): By gender, females spent $49.18 (control) vs. $51.30 (treatment), and males spent $61.82 (control) vs. $70.83 (treatment).

Chart (b): By age, customers age ≤ 35 spent $56.68 (control) vs. $43.41 (treatment); ages 35–55 spent $59.21 (control) vs. $71.19 (treatment); ages > 55 spent $40.24 (control) vs. $79.07 (treatment).}
    \label{fig:all-bars}
    \vspace{-1\baselineskip}
\end{figure*}

\begin{table}[t]
\footnotesize
\begin{booktabs}{
    colspec={ccX[c]c},
    width=\linewidth
}
\toprule
\textbf{Participant} & \textbf{Gender} &  \textbf{Job Title} & \textbf{Team}   \\ \midrule
P1  & F  & Project Manager       & PM Team      \\
P2  & M    & Software Development Manager       & Engineer Team     \\
P3  & F    & Product Manager       & PM Team     \\
P4  & M    & Product Manager       & PM Team  \\
P5  & M   & Product Manager       & PM Team  \\
P6  & M  & Machine Learning Researcher & Science Team   \\ 

\bottomrule
\end{booktabs}%
\caption{Demographics of formative study participants. }
\label{tab:participants}
\vspace{-1\baselineskip}
\end{table}

\section{Example of Persona Generated by \systemname}
\label{app:example}
\begin{lstlisting}
Persona: Marcus

Background:
Marcus is a 35-year-old freelance graphic designer living in Austin, Texas. After working for a decade in various creative agencies, he transitioned to freelancing to gain more control over his schedule and focus on passion projects, such as illustrating indie game assets and creating digital art for local musicians.

Demographics:

Age: 35

Gender: Male

Education: Bachelor's degree in Visual Communication

Profession: Freelance Graphic Designer

Income: $70,000 (variable based on projects)

Financial Situation:
Marcus earns a decent living from his freelance gigs, though his income can fluctuate. He's financially stable but cautious about big expenses. He sets aside part of his earnings for travel and software upgrades, which are essential for his work.

Shopping Habits:
Marcus enjoys discovering unique or niche products, especially tech gadgets, art supplies, and streetwear. He prefers shopping online for the variety and reads reviews carefully. He's brand-loyal when it comes to tools he relies on, like his drawing tablet and design software. For clothing, he's drawn to bold, graphic-heavy items that reflect his artistic vibe.

Professional Life:
Marcus works from a home studio that doubles as a creative space. He collaborates remotely with clients from different industries, juggles multiple deadlines, and often pulls late nights. He frequently updates his portfolio and maintains a strong social media presence to attract new clients.

Personal Style:
Marcus has an edgy and expressive fashion sense. He wears large-sized clothing and gravitates toward dark tones with pops of neon or graphic prints. Comfort is important, but he likes to make a visual statement with what he wears. His go-to outfit is a soft hoodie with a custom design, black joggers, and high-top sneakers.
\end{lstlisting}

\section{Persona Generation Prompt}
{%
\begin{lstlisting}
You are a helpful assistant that generates diverse personas.
Examples:
[PERSONA EXAMPLE]

Generate a persona using the above examples. The persona should be different from previous personas to ensure diversity. The persona should:
- Have the age of {age}
- Be {gender}
- Have an income between ${income_range[0]} and ${income_range[1]}
Provide the persona in the same format as the examples.
Only output the persona, no other text.
\end{lstlisting}
}

\section{Interview Protocol}

\revise{1. Overall Experimentation Practice \\
Can you describe your general workflow for running A/B testing experiments in your company? \\
2. End-to-End Workflow and Timing \\
How does an experiment progress from an initial design idea to a final deployment decision, and how long do the major stages typically take? \\
3. Pain Points and Challenges \\
What are the main challenges or friction points you encounter when designing, running experiments or analyzing results? \\
4. Experimentation Tools and Platform Evaluation \\
How well does the existing A/B testing experiment tool support your work, and what new features would you want? \\
5. Other \\
Is there anything else you would like to add, or do you have any questions for me?}

\end{document}